\documentclass[conference]{IEEEtran}

\usepackage{amsmath,amssymb,mathtools}
\usepackage{bm}
\usepackage{cite}
\usepackage{graphicx}
\usepackage{booktabs}
\usepackage{xcolor}
\usepackage{soul}
\sethlcolor{yellow}
\usepackage{url}

\begin{document}

\title{Beam Squint and Aperture--Bandwidth Limitations in Wideband RIS-Assisted THz Links}

\author{
\IEEEauthorblockN{Waqas Khalid, Chiew Foong Kwong, David Chieng, and Qianyu Liu}
\IEEEauthorblockA{
Department of Electrical and Electronic Engineering, Next Generation Internet of Everything Laboratory (NGIoE Lab)\\
University of Nottingham Ningbo China, Ningbo, 315100, China\\
Email: Waqas.Khalid@nottingham.edu.cn, Chiew-Foong.Kwong@nottingham.edu.cn, \\ David.Chieng@nottingham.edu.cn, Qianyu.Liu@nottingham.edu.cn}
}

\maketitle

\begin{abstract}
This paper analyzes the ergodic rate of a wideband reconfigurable
intelligent surface (RIS)-assisted terahertz link under beam squint and
finite-resolution phase control. Independent Rayleigh fading on the two
RIS-assisted hops produces double-Rayleigh cascaded amplitudes. Exact
second- and fourth-order channel moments are derived by incorporating
the frequency-dependent array response and phase-quantization errors.
Moment matching then provides a Gamma approximation for the received
power on each subcarrier and a tractable closed-form approximation for
the wideband ergodic rate. A first-null bandwidth approximation
characterizes the aperture--bandwidth limitation of frequency-flat RIS
control. Monte Carlo simulations validate the analytical framework.
\end{abstract}

\begin{IEEEkeywords}
RIS, terahertz communications, ergodic rate, beam squint, phase
quantization.
\end{IEEEkeywords}

\section{Introduction}
\label{sec:introduction}

Terahertz (THz) communication can support extremely high data rates
because of the abundant spectrum available at submillimeter wavelengths.
Nevertheless, THz links experience severe spreading loss, molecular
absorption, and sensitivity to blockage, which restrict their coverage
\cite{akyildiz2014terahertz}. Reconfigurable intelligent surfaces
(RISs) can alleviate these limitations by creating controllable
reflected paths without conventional radio-frequency chains
\cite{diRenzo2020smart}. The large number of reflecting elements available at THz frequencies can also provide considerable passive beamforming gain. RIS research has
further considered simultaneous transmitting and reflecting
architectures, active operation, hardware impairments, phase-dependent
amplitude responses, physical-layer security, and malicious-surface
threats
\cite{khalid2023star,khalid2025active,khalid2022rateenergy,
khalid2021fdjamming,khalid2024pls,khalid2025malicious}.

Most statistical analyses of RIS-assisted links adopt a narrowband
model in which one RIS phase configuration coherently combines the
reflected signals across the operating band
\cite{khalid2023star,khalid2025active,khalid2021fdjamming,
khalid2024pls,khalid2025malicious}. This assumption becomes inaccurate
for wideband THz transmission. Practical frequency-flat phase shifts
designed at the center frequency cannot perfectly align the reflected
components on all subcarriers. The resulting frequency-dependent phase
progression, known as beam squint, reduces coherent array gain and
becomes more severe as the bandwidth or RIS aperture increases
\cite{yan2022wideband,qian2024wideband}.

Existing wideband RIS studies have mainly addressed beamforming and
phase-shift optimization under deterministic or geometry-based channel
models. These studies condition on a channel realization and
optimize an instantaneous objective, whereas statistical performance
analysis averages over fading to derive channel distributions or
long-term metrics. The present work follows the latter direction: its
RIS profile is prescribed by center-frequency alignment, and its main
contribution is the derivation of channel-power moments and ergodic-rate
expressions rather than a new beamforming algorithm. Stochastic
performance analyses have considered fading distributions, phase
imperfections, and pointing errors
\cite{le2023risthz,mosleh2023ergodic,totaKhel2023secrecy,
durgada2022risthz}. However, these research directions do not jointly
characterize cascaded small-scale fading, frequency-dependent beam
squint, and discrete RIS phases. The analytical difficulty is that the
effective channel becomes a frequency-dependent complex sum of
double-Rayleigh amplitudes with deterministic beam-squint phases and
random phase-quantization errors, whose received-power distribution has
no convenient exact form.

Accordingly, this paper develops a tractable ergodic-rate analysis for
a wideband RIS-assisted THz link. Its main contributions are as follows:
\begin{itemize}
    \item Exact second- and fourth-order moments of the effective
    channel are derived under cascaded Rayleigh fading, beam squint, and
    finite-resolution phase control. The expressions explicitly
    incorporate the subcarrier-dependent finite-aperture array factors.

    \item The received power is approximated by a moment-matched Gamma
    random variable, which yields a closed-form approximation for the
    per-subcarrier and wideband ergodic rates.

    \item A first-null bandwidth condition is derived to reveal how RIS
    size and propagation geometry limit the bandwidth supported by a
    frequency-flat phase configuration. Monte Carlo simulations
    validate the analytical framework.
\end{itemize}

\section{System and Channel Model}
\label{sec:system_model}

\begin{figure}[t]
    \centering
    \includegraphics[
        width=\columnwidth,
        keepaspectratio
    ]{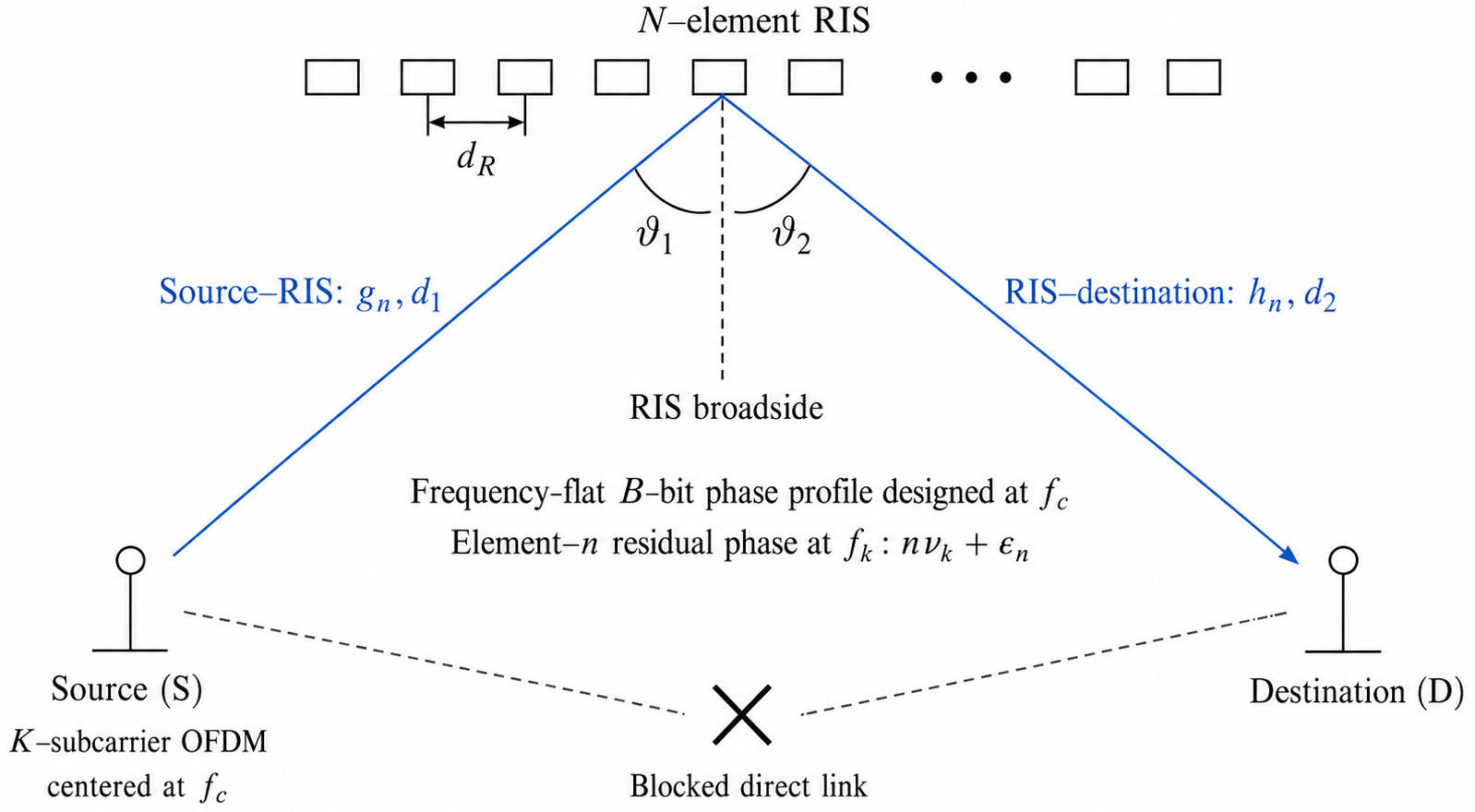}
    \caption{Wideband SISO THz link assisted by an $N$-element
    frequency-flat RIS. The common $B$-bit phase profile designed at
    $f_{\mathrm c}$ produces residual phase
    $n\nu_k+\varepsilon_n$ on subcarrier $k$, while the direct link is
    blocked.}
    \label{fig:system_model}
\end{figure}

Consider the wideband single-input single-output THz system illustrated
in Fig.~\ref{fig:system_model}, where a source communicates with a
destination through an $N$-element passive RIS using a frequency-flat
phase profile. The direct link is unavailable because of severe
blockage. The RIS is modeled as a uniform linear array with
inter-element spacing $d_{\mathrm R}$. For a $K$-subcarrier
orthogonal frequency-division multiplexing waveform, the frequency of
subcarrier $k$ is
\begin{equation}
f_k=f_{\mathrm c}
+\left(k-\frac{K-1}{2}\right)\Delta f,
\quad k=0,\ldots,K-1,
\label{eq:subcarrier_frequency}
\end{equation}
where $f_{\mathrm c}$ is the center frequency, $\Delta f$ is the
subcarrier spacing, and $W=K\Delta f$ is the occupied bandwidth. We
assume odd $K$ and define the center-subcarrier index as
$k_{\mathrm c}\triangleq(K-1)/2$.

Let $s_k$ be the unit-power symbol transmitted on subcarrier $k$ with
power $p_k$. The received signal is
\begin{equation}
y_k=\sqrt{p_k}\,H_k s_k+n_k,
\label{eq:received_signal}
\end{equation}
where $H_k$ is the RIS-assisted channel and
$n_k\sim\mathcal{CN}(0,\sigma_k^2)$ is additive Gaussian noise. The
powers satisfy $\sum_{k=0}^{K-1}p_k\leq P$, and
$\sigma_k^2=N_0\Delta f$, where $P$ and $N_0$ denote the total
transmit-power budget and noise power spectral density, respectively.
Unless otherwise stated, equal allocation, $p_k=P/K$, is assumed. The
normalized comparisons in Section~\ref{sec:results} impose
$\rho_k=\rho_{k_{\mathrm c}}$ to isolate beam squint.

Let $d_1$ and $d_2$ denote the source--RIS and RIS--destination
distances. The deterministic cascaded THz amplitude coefficient is
\begin{equation}
\beta_k=
\eta_k
\left(\frac{c}{4\pi f_kd_1}\right)
\left(\frac{c}{4\pi f_kd_2}\right)
\exp\left[
-\frac{\kappa_{\mathrm{abs}}(f_k)(d_1+d_2)}{2}
\right],
\label{eq:thz_path_gain}
\end{equation}
where $c$ is the speed of light,
$\kappa_{\mathrm{abs}}(f_k)$ is the molecular absorption coefficient,
and $\eta_k$ captures the antenna gains, RIS element response, and
reflection efficiency \cite{akyildiz2014terahertz,le2023risthz}.

The normalized small-scale coefficients of the two RIS hops satisfy $g_n\sim\mathcal{CN}(0,\Omega_g)$, and $h_n\sim\mathcal{CN}(0,\Omega_h)$, and are independent across elements and between hops \cite{khalid2019sensing}. Within each
realization, their amplitudes are assumed constant over the OFDM band,
while the frequency-dependent inter-element phase is retained through
$\nu_k$. The single-tap assumption isolates beam squint from multipath
frequency selectivity. Additional delayed components in practical
THz channels would make the hops frequency selective and alter the
received-power moments, potentially increasing the Gamma-model
mismatch. Extending the analysis would require the tap powers, delays,
and cross terms; hence, the present model is most accurate for sparse
or dominant-path THz channels.

Assuming perfect knowledge of the cascaded channel phases at
$f_{\mathrm c}$, the RIS aligns the reflected contributions at the
center frequency; channel-estimation and control overhead are outside
the present scope. With $B$-bit phase control, the residual
quantization error is modeled as
$\varepsilon_n\sim\mathcal{U}[-\Delta_B,\Delta_B]$, where
$\Delta_B=\pi/2^B$ \cite{totaKhel2023secrecy}. The errors are mutually
independent and independent of the fading coefficients.

Because the same RIS profile is applied to every subcarrier, phase
alignment is generally lost when $f_k\neq f_{\mathrm c}$. After
center-frequency phase compensation, the effective channel is
\begin{equation}
H_k=\beta_k
\sum_{n=0}^{N-1}
X_n\exp\left[j\left(n\nu_k+\varepsilon_n\right)\right],
\qquad
X_n=|g_n||h_n|,
\label{eq:effective_channel}
\end{equation}
where the frequency-dependent phase progression is
\begin{equation}
\nu_k=
\frac{2\pi(f_k-f_{\mathrm c})d_{\mathrm R}}{c}
\left(\sin\vartheta_1+\sin\vartheta_2\right),
\label{eq:beam_squint_phase}
\end{equation}
and $\vartheta_1$ and $\vartheta_2$ are the signed incident and
reflected angles measured from the RIS broadside. Under the adopted
array convention, the cascaded inter-element delay is
$d_{\mathrm R}(\sin\vartheta_1+\sin\vartheta_2)/c$. Thus,
$\nu_k=0$ at the center frequency and generally increases in magnitude
toward the band edges \cite{yan2022wideband,qian2024wideband}.

For subsequent analysis, define
\begin{equation}
Z_k=
\left|
\sum_{n=0}^{N-1}
X_n e^{j(n\nu_k+\varepsilon_n)}
\right|^2,
\quad
\gamma_k=\rho_kZ_k,
\quad
\rho_k=\frac{p_k|\beta_k|^2}{\sigma_k^2}.
\label{eq:snr_definition}
\end{equation}
The wideband ergodic rate is
\begin{equation}
\overline{R}
=
\frac{1}{K}
\sum_{k=0}^{K-1}
\mathbb{E}\!\left[
\log_2(1+\rho_kZ_k)
\right].
\label{eq:wideband_rate}
\end{equation}
A tractable statistical characterization of $Z_k$ is developed in the following section.

\section{Statistical and Ergodic-Rate Analysis}
\label{sec:analysis}
\subsection{Effective-Channel Moments}
\label{subsec:channel_moments}

For convenience, write
\begin{equation}
S_k=\sum_{n=0}^{N-1}
X_n e^{j(n\nu_k+\varepsilon_n)},
\qquad
Z_k=|S_k|^2.
\end{equation}
Since $X_n=|g_n||h_n|$ is the product of two independent Rayleigh
amplitudes, it follows a double-Rayleigh cascaded distribution
\cite{chien2021coverage}. Its $r$th-order moment is
\begin{equation}
\mu_r
\triangleq
\mathbb{E}[X_n^r]
=
(\Omega_g\Omega_h)^{r/2}
\Gamma^2\left(1+\frac{r}{2}\right),
\quad r=1,\ldots,4.
\label{eq:double_rayleigh_moments}
\end{equation}
The characteristic coefficients of the phase-quantization error are
\begin{equation}
q_\ell
\triangleq
\mathbb{E}[e^{j\ell\varepsilon_n}]
=
\frac{\sin(\ell\Delta_B)}{\ell\Delta_B},
\qquad \ell\in\{1,2\}.
\label{eq:phase_error_moments}
\end{equation}
Both coefficients approach one as $B$ increases; for continuous-phase
control, $q_1=q_2=1$ by continuity.

The frequency-dependent finite-aperture array factors are
\begin{equation}
\begin{aligned}
D_{\ell,k}
&\triangleq
\sum_{n=0}^{N-1}e^{j\ell n\nu_k}\\
&=
e^{j\ell(N-1)\nu_k/2}
\frac{\sin(N\ell\nu_k/2)}
     {\sin(\ell\nu_k/2)}.
\end{aligned}
\label{eq:array_factor}
\end{equation}
When $\ell\nu_k=2\pi m$, $m\in\mathbb{Z}$, the ratio is interpreted
by continuity and $D_{\ell,k}=N$. Thus,
$D_{1,k_{\mathrm c}}=D_{2,k_{\mathrm c}}=N$, while destructive
combining may occur away from the center frequency.

For compactness, define
\begin{equation}
\begin{aligned}
\alpha&=\mu_1q_1,\quad
\xi=\mu_2-\alpha^2,\quad
\zeta=\mu_2q_2-\alpha^2,\\
t&=\mu_3q_1-\alpha\mu_2q_2
   -2\alpha\mu_2+2\alpha^3,\\
u&=\mu_4-4\alpha\mu_3q_1
   +2\alpha^2\mu_2q_2\\
&\quad+4\alpha^2\mu_2-3\alpha^4.
\end{aligned}
\label{eq:central_moment_terms}
\end{equation}
These quantities represent the required central mixed moments of
$X_ne^{j\varepsilon_n}$.

\noindent\textbf{Lemma 1:}
The first and second moments of $Z_k$ are
\begin{equation}
\begin{aligned}
M_{1,k}
&\triangleq \mathbb{E}[Z_k]\\
&=N\mu_2+\alpha^2\bigl(|D_{1,k}|^2-N\bigr),
\end{aligned}
\label{eq:first_power_moment}
\end{equation}
and
\begin{equation}
\begin{aligned}
M_{2,k}
&\triangleq \mathbb{E}[Z_k^2]\\
&=
\alpha^4|D_{1,k}|^4
+4N\alpha^2\xi|D_{1,k}|^2\\
&\quad
+2\alpha^2\zeta
\operatorname{Re}\!\left\{
D_{2,k}(D_{1,k}^{*})^2
\right\}\\
&\quad
+4\alpha t|D_{1,k}|^2
+Nu\\
&\quad
+\zeta^2\left(|D_{2,k}|^2-N\right)
+2N(N-1)\xi^2.
\end{aligned}
\label{eq:second_power_moment}
\end{equation}

\emph{Proof:}
Let $Y_n=X_ne^{j\varepsilon_n}=\alpha+\widetilde{Y}_n$, where
$\mathbb{E}[\widetilde{Y}_n]=0$. Substitution into
$S_k=\sum_n e^{jn\nu_k}Y_n$ and collection of equal- and
distinct-index terms yield \eqref{eq:first_power_moment}. For
\eqref{eq:second_power_moment}, use
$\mathbb{E}[|\widetilde{Y}_n|^2]=\xi$,
$\mathbb{E}[\widetilde{Y}_n^2]=\zeta$,
$\mathbb{E}[\widetilde{Y}_n^2\widetilde{Y}_n^*]=t$, and
$\mathbb{E}[|\widetilde{Y}_n|^4]=u$, and collect the resulting sums
through $D_{1,k}$ and $D_{2,k}$. \hfill$\blacksquare$

The exact distribution of $Z_k$ is intractable because it is the
squared magnitude of a frequency-dependent sum of double-Rayleigh
variables. Motivated by related RIS channel approximations
\cite{chien2021coverage,totaKhel2023secrecy}, we use
\begin{equation}
\begin{aligned}
Z_k&\ \dot{\sim}\
\operatorname{Gamma}(\kappa_k,\theta_k),\\
\kappa_k
&=
\frac{M_{1,k}^2}
     {M_{2,k}-M_{1,k}^2},\\
\theta_k
&=
\frac{M_{2,k}-M_{1,k}^2}
     {M_{1,k}},
\end{aligned}
\label{eq:gamma_parameters}
\end{equation}
where $\kappa_k$ and $\theta_k$ are the shape and scale parameters.
The resulting PDF is
\begin{equation}
f_{Z_k}(z)
\approx
\frac{z^{\kappa_k-1}}
{\Gamma(\kappa_k)\theta_k^{\kappa_k}}
\exp\left(-\frac{z}{\theta_k}\right),
\qquad z\geq0.
\label{eq:gamma_pdf}
\end{equation}
Accordingly,
$F_{Z_k}(z)\approx P(\kappa_k,z/\theta_k)$, where
$P(a,x)=\gamma(a,x)/\Gamma(a)$ is the regularized lower incomplete
Gamma function. Both Gamma parameters vary across subcarriers through
$D_{1,k}$ and $D_{2,k}$, thereby retaining the statistical effect of
beam squint.

\subsection{Closed-Form Ergodic-Rate Analysis}
\label{subsec:ergodic_rate}

Using \eqref{eq:gamma_pdf}, the per-subcarrier ergodic rate admits the
following closed-form approximation.

\noindent\textbf{Proposition 1:}
\begin{equation}
\overline{R}_k
\approx
\frac{1}{\Gamma(\kappa_k)\ln 2}
\mathrm{G}_{2,3}^{3,1}\!
\left(
\frac{1}{\rho_k\theta_k}
\;\middle|\;
\begin{matrix}
0,1\\
0,0,\kappa_k
\end{matrix}
\right),
\label{eq:closed_form_rate}
\end{equation}
where $\mathrm{G}_{p,q}^{m,n}(\cdot)$ denotes the Meijer-$G$
function. The wideband rate follows by substituting
\eqref{eq:closed_form_rate} into \eqref{eq:wideband_rate}.

\emph{Proof:}
Substituting \eqref{eq:gamma_pdf} into the rate expectation and setting
$x=z/\theta_k$ gives
\begin{equation}
\overline{R}_k
\approx
\frac{1}{\Gamma(\kappa_k)\ln 2}
\int_{0}^{\infty}
x^{\kappa_k-1}e^{-x}
\ln(1+\rho_k\theta_kx)\,\mathrm{d}x.
\end{equation}
Expressing the logarithm in Meijer-$G$ form and applying the standard
Mellin-convolution identity \cite{gradshteyn2014table} gives
\eqref{eq:closed_form_rate}. \hfill$\blacksquare$

Equation~\eqref{eq:closed_form_rate} avoids Monte Carlo averaging and
requires only standard special functions. If a symbolic Meijer-$G$
routine is unavailable, the preceding integral can be evaluated using
Gauss--Laguerre quadrature.

\noindent\textit{Aperture--bandwidth implication:}
At the band edge, $|f_k-f_{\mathrm c}|\simeq W/2$, and the first null
of \eqref{eq:array_factor} occurs when $N|\nu_k|/2=\pi$. Therefore,
\begin{equation}
W_{\mathrm{null}}
\simeq
\frac{2c}
{N d_{\mathrm R}
\left|\sin\vartheta_1+\sin\vartheta_2\right|}.
\label{eq:first_null_bandwidth}
\end{equation}
Thus, $W<W_{\mathrm{null}}$ keeps the operating band within the
principal coherent lobe. Increasing the RIS aperture improves the
center-frequency gain but reduces the bandwidth supported by one
frequency-flat phase configuration. When
$\sin\vartheta_1+\sin\vartheta_2=0$, the linear phase progression
vanishes and $W_{\mathrm{null}}\rightarrow\infty$.

\section{Numerical Results}
\label{sec:results}

Monte Carlo simulations validate the proposed analysis. Unless otherwise
stated, $f_{\mathrm c}=0.3~\mathrm{THz}$,
$\lambda_{\mathrm c}=c/f_{\mathrm c}$, $K=129$,
$d_{\mathrm R}=\lambda_{\mathrm c}/2$, $N=64$, $B=2$, and
$W=30~\mathrm{GHz}$. The average channel powers are
$\Omega_g=\Omega_h=1$, with
$\vartheta_1=\vartheta_2=10^\circ$. Each empirical CDF uses $10^6$
independent realizations of the double-Rayleigh channels and
phase-quantization errors.

\begin{figure*}[t]
    \centering
    \includegraphics[width=\textwidth]{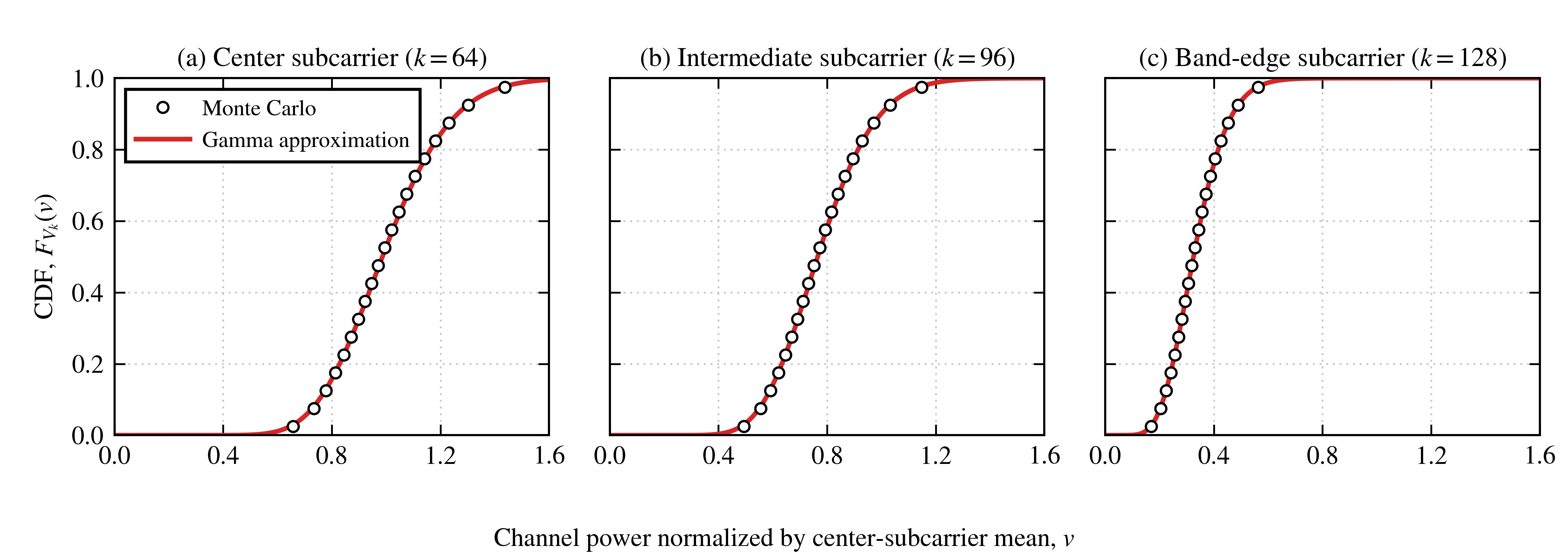}
    \caption{Empirical and Gamma-approximated CDFs of
    $V_k=Z_k/\mathbb{E}[Z_{k_{\mathrm c}}]$ at the center
    ($k=64$), intermediate ($k=96$), and upper band-edge
    ($k=128$) subcarriers.}
    \label{fig:gamma_validation}
\end{figure*}

\begin{figure}[t]
    \centering
    \includegraphics[
        width=\columnwidth,
        keepaspectratio
    ]{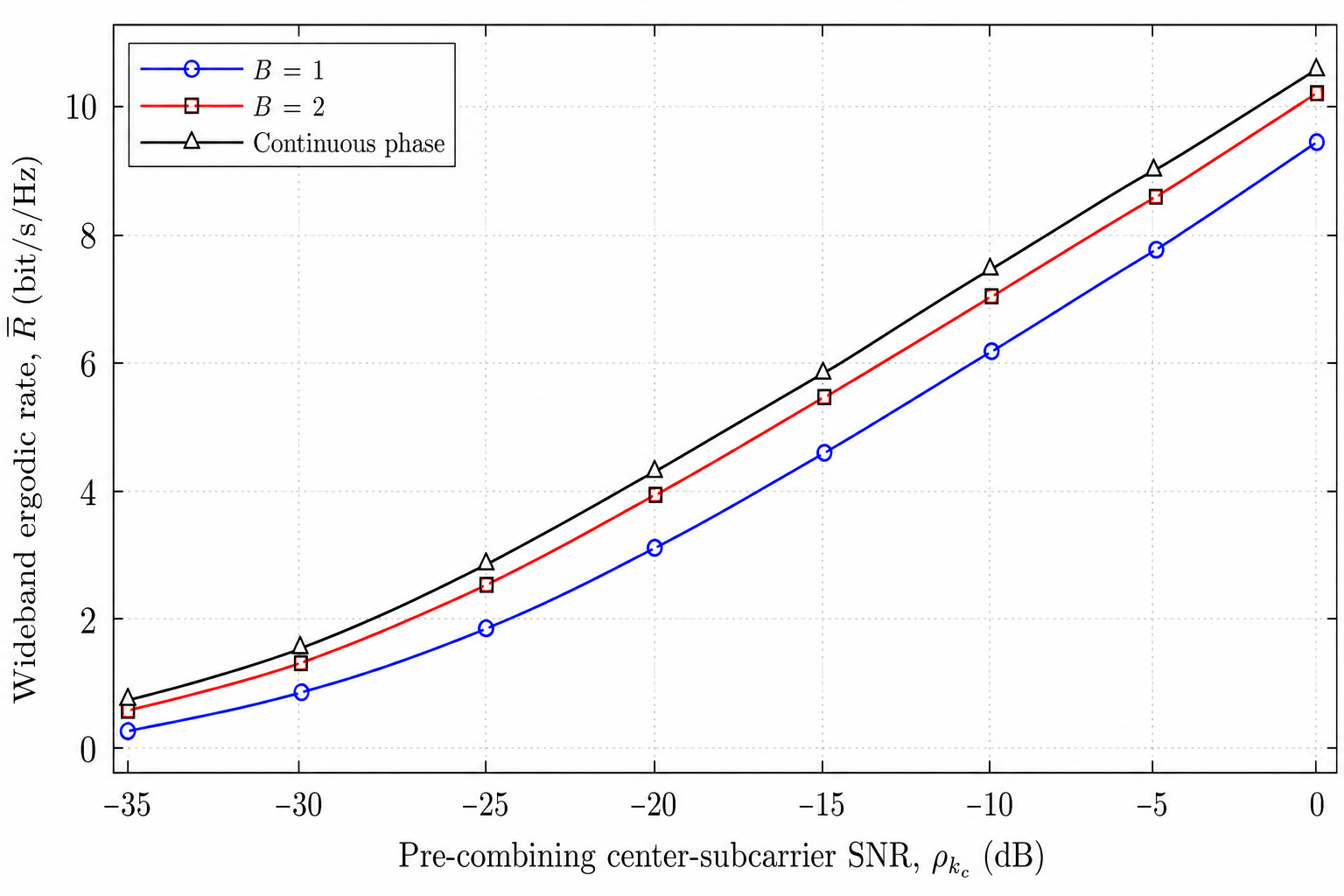}
    \caption{Wideband ergodic rate versus the nominal center-subcarrier
    pre-combining SNR for different RIS phase resolutions. Lines denote
    analysis and markers denote Monte Carlo simulation.}
    \label{fig:rate_snr}
\end{figure}

\begin{figure}[t]
    \centering
    \includegraphics[
        width=\columnwidth,
        keepaspectratio
    ]{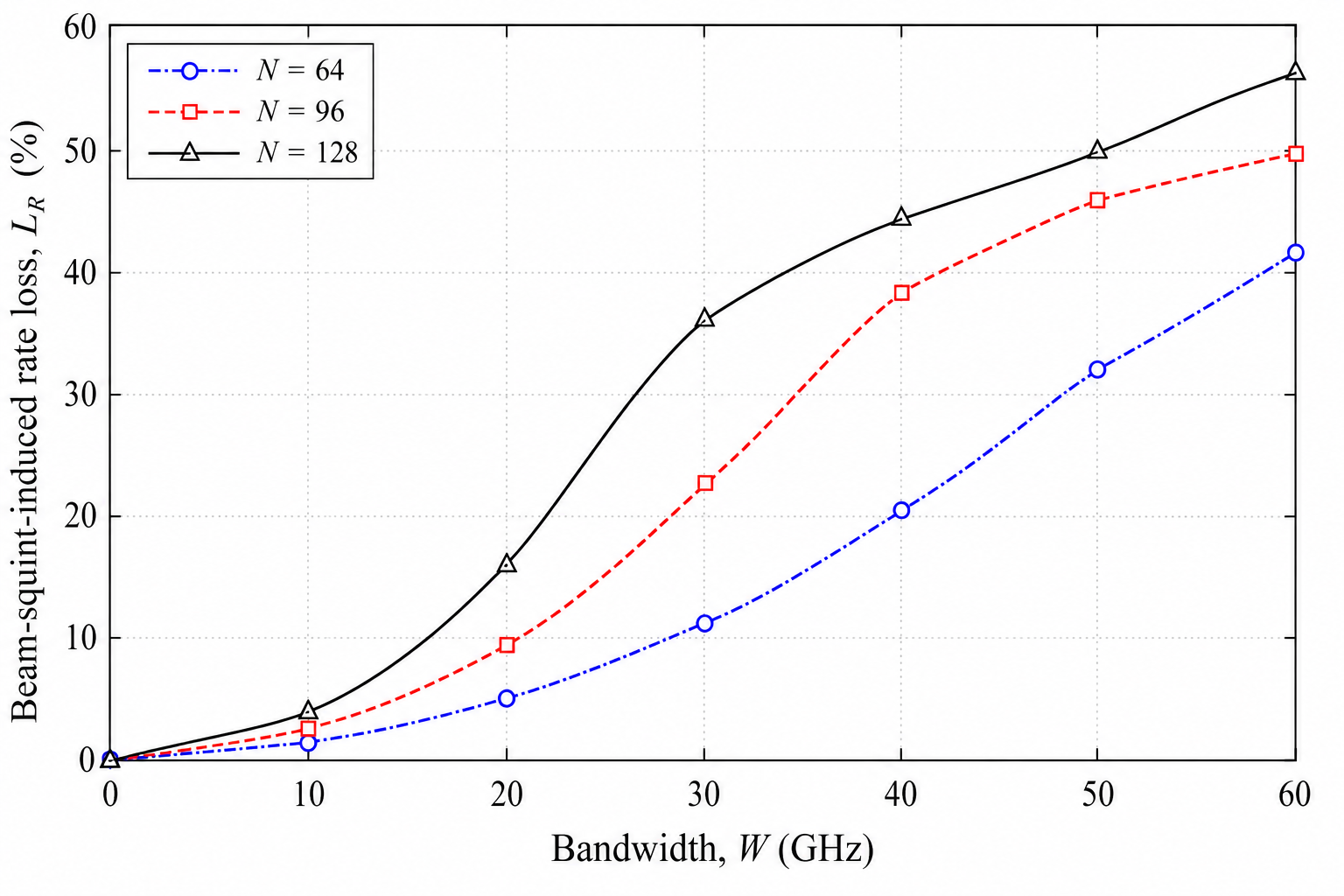}
    \caption{Beam-squint-induced wideband rate loss of frequency-flat
    RIS control relative to subcarrier-dependent phase alignment.
    Lines denote analysis and markers denote Monte Carlo simulation.}
    \label{fig:rate_bandwidth}
\end{figure}

\subsection{Accuracy of the Gamma Approximation}

Fig.~\ref{fig:gamma_validation} compares the empirical and
Gamma-approximated CDFs at $k\in\{64,96,128\}$, representing the
center, intermediate, and upper band-edge subcarriers. To preserve the
frequency-dependent power variation, all subcarriers are normalized by
the center-subcarrier mean:
$V_k=Z_k/M_{1,k_{\mathrm c}}$, where $k_{\mathrm c}=64$. Hence,
\begin{equation}
F^\Gamma_{V_k}(v)
\triangleq
P\!\left(
\kappa_k,
\frac{vM_{1,k_{\mathrm c}}}{\theta_k}
\right).
\end{equation}
The analytical curves closely follow the Monte Carlo markers. The
normalized mean powers decrease from $1$ at $k=64$ to $0.7772$ and
$0.3352$ at $k=96$ and $128$, producing the progressive leftward CDF
shift. Meanwhile, $\kappa_k$ decreases from $24.9984$ to $21.3863$
and $10.9913$, indicating increased relative dispersion toward the
band edge.

The distributional mismatch is quantified using the
Kolmogorov--Smirnov (KS) distance
\begin{equation}
D_{\mathrm{KS},k}\triangleq
\sup_{v\geq0}
\left|
\widehat{F}^{\mathrm{MC}}_{V_k}(v)
-
F^{\Gamma}_{V_k}(v)
\right|.
\end{equation}
The KS distances are 0.0077, 0.0081, and 0.0044 for subcarriers
64, 96, and 128, respectively. Their values below 0.01 quantitatively
confirm the close agreement shown in
Fig.~\ref{fig:gamma_validation}. Thus, the Gamma model captures both
the power loss and distributional change induced by beam squint.

\subsection{Wideband Ergodic Rate and Phase Resolution}

Fig.~\ref{fig:rate_snr} shows the wideband ergodic rate versus the
nominal center-subcarrier pre-combining SNR,
$\rho_{k_{\mathrm c}}$. We impose
$\rho_k=\rho_{k_{\mathrm c}}$ on every subcarrier to isolate beam
squint and phase quantization from frequency-dependent path loss. The
continuous-phase benchmark eliminates quantization errors but retains
the frequency-flat profile designed at $f_{\mathrm c}$ and therefore
remains affected by beam squint.

The analytical curves obtained from \eqref{eq:closed_form_rate} closely
match the Monte Carlo markers over the entire SNR range. At
$\rho_{k_{\mathrm c}}=-20~\mathrm{dB}$, continuous-, two-, and one-bit
phase control achieve approximately $4.21$, $3.94$, and
$3.06~\mathrm{bit/s/Hz}$, respectively. Thus, two-bit control improves
the rate by $28.8\%$ over one-bit control and remains only $6.6\%$
below continuous-phase control. At $0~\mathrm{dB}$, the corresponding
rates are $10.77$, $10.47$, and $9.49~\mathrm{bit/s/Hz}$. Increasing
the resolution beyond two bits therefore provides only a modest gain
under the considered conditions.

\subsection{Aperture--Bandwidth Rate Loss}

Fig.~\ref{fig:rate_bandwidth} examines the bandwidth sensitivity of
frequency-flat RIS control for different RIS sizes. The ideal benchmark
independently aligns the RIS phases on every subcarrier and serves only
as an upper reference. Both schemes use identical phase resolution,
nominal SNR, geometry, and channel parameters. The beam-squint-induced
relative rate loss is
\begin{equation}
L_R(W)
=
100\left[
1-
\frac{\overline{R}_{\mathrm{flat}}(W)}
     {\overline{R}_{\mathrm{ideal}}(W)}
\right]\%.
\end{equation}
We impose $\rho_k=\rho_{k_{\mathrm c}}=-20~\mathrm{dB}$ on all
subcarriers; thus, the comparison isolates beam squint and is not a
fixed-total-power bandwidth sweep.

The loss approaches zero in the narrowband limit and increases with
bandwidth and RIS size. At $W=30~\mathrm{GHz}$, the losses are
approximately $11.0\%$, $22.2\%$, and $36.0\%$ for $N=64$, $96$,
and $128$, respectively. Equation~\eqref{eq:first_null_bandwidth}
predicts first-null bandwidths of approximately $54$, $36$, and
$27~\mathrm{GHz}$. Therefore, at $W=30~\mathrm{GHz}$, the band edge
has crossed the first null for $N=128$, but not for the smaller RISs.
At $W=60~\mathrm{GHz}$, the losses reach $41.4\%$, $49.7\%$, and
$56.2\%$, respectively.

No discontinuity occurs at $W_{\mathrm{null}}$ because
$\overline{R}$ averages all subcarrier rates, whereas
\eqref{eq:first_null_bandwidth} concerns only the band-edge
subcarrier. These results confirm the inverse aperture--bandwidth
relationship. A larger relative loss indicates stronger beam-squint
sensitivity, but not necessarily a smaller absolute rate.

\vspace{-2mm}
\section{Conclusion}
\label{sec:conclusion}

This paper developed a tractable ergodic-rate analysis for wideband
RIS-assisted THz links under cascaded Rayleigh fading, beam squint, and
finite-resolution phase control. Exact second- and fourth-order channel
moments were derived using frequency-dependent finite-aperture factors
and phase-quantization errors. Moment matching then yielded a Gamma
approximation and a closed-form wideband ergodic-rate expression.
Monte Carlo results validated the proposed distribution and rate
analysis. They also showed that two-bit control captures most of the
continuous-phase gain, whereas beam-squint-induced loss grows with
bandwidth and RIS size. The derived first-null condition explains the
tradeoff between center-frequency array gain and wideband coherent
combining.

Although analytical, the framework is validated only through Monte
Carlo simulations under single-tap Rayleigh fading, perfect
center-frequency channel knowledge, and idealized phase-quantization
errors. Measurement-based validation is not provided, and the accuracy
may change under practical multipath, channel-estimation errors, and
hardware impairments. Future work will consider measured multipath THz
channels, hardware experiments, planar RIS geometries, and
frequency-dependent element responses.


\bibliographystyle{IEEEtran}
\bibliography{references}

\end{document}